\documentclass[twocolumn,letter]{jpsj2}

\title{
Theory for Coupled SDW and Superconducting Order in FFLO State of CeCoIn$_5$
}
\author{
Kazumasa {\sc Miyake}$^{1,2}$
}
\inst
{$^{1}$
Division of Materials Physics, Department of Materials Engineering Science \\
Graduate School of Engineering Science, Osaka University, Toyonaka 560-8531,\\
$^{2}$
Commissariat \`a l' \'Energie Atomic, INAC, SPSMS, 17 rue des Martyrs, 
38054 Grenoble, France
}

\recdate
{
September 12, 2008
}

\abst
{
The mechanism of incommensurate (IC) spin-density-wave (SDW) order observed 
in the Fulde-Ferrell-Larkin-Ovchinnikov (FFLO) phase of CeCoIn$_5$ 
is discussed on the basis of new mode-coupling scheme among IC-SDW order, 
two superconducting orders of FFLO with B$_{1{\rm g}}$ 
($d_{x^{2}-y^{2}}$) symmetry and $\pi$-pairing of odd-parity.  
Unlike the mode-coupling schemes proposed 
by Kenzelmann et al, Sciencexpress, 21 August (2008), that proposed 
in the present Letter can offer a simple explanation for why the IC-SDW 
order is observed only in FFLO phase and the IC wave vector is rather 
robust against the magnetic field.  
}

\kword{
FFLO phase in CeCoIn$_5$, 
incommensurate spin-density-wave order, 
}
\begin{document}
\sloppy
\maketitle
Recent discovery of incommensurate (IC) spin-density-wave (SDW) order 
in the Fulde-Ferrell-Larkin-Ovchinnikov (FFLO) phase of CeCoIn$_5$ 
has attracted intense attention~\cite{Kenzelmann}.  This result is 
consistent with that of NMR Knight shift measurements suggesting the 
existence of field-induced magnetism~\cite{Young}.  
Kenzelmann et al reported that an IC-SDW order in the $c$-direction 
with the wave vector 
${\bf Q}=(h\pi/a,h\pi/a,0.5\pi/c)$, $h=0.56$ and $a$ being the lattice 
constant in the $ab$-plane, exists in the so-called FFLO phase 
under the magnetic field $H$ along the (1,1,0) direction.  They also 
claimed that the wave vector ${\bf Q}$ is essentially constant against the 
magnetic field strength $H$ within an experimental resolution.  In order 
to explain their results, they proposed a scenario on the basis of 
mode-couplings among the IC-SDW order $M_{q}$, $d$-wave pairing 
$\Delta_{d}$, additional superconducting orders with finite center-of-mass 
momentum $\Delta^{i}_{n,-q}$, with different combination of $(i,n)$ 
on the symmetry requirement.  While the scenario explains the appearance 
of the IC-SDW order, it fails to explain the reason why the IC-SDW is 
observed only in the FFLO phase and the ordering wave vector ${\bf Q}$ 
is robust against the magnetic field.  

\begin{figure}[h]
\begin{center}
\rotatebox{0}{\includegraphics[width=0.6\linewidth]{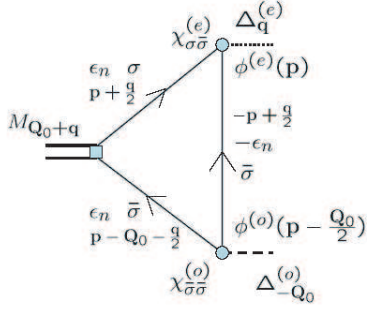}}
\caption{Feynman diagram for the mode-coupling term eq.(\ref{eq:1}) among 
IC-SDW magnetization, $M_{{\bf Q}_0+{\bf q}}$, $d$-wave SC gap of FFLO state, 
$\Delta_{\bf q}$, and the so-called $\pi$-gap of odd-parity 
with center-of mass momentum $Q_{0}$, $\Delta_{-{\bf Q}_0}$.  
The solid line presents the Green function of the quasiparticles in the 
normal state.  
}
\label{Fig:1}
\end{center}
\end{figure}

The purpose of this Letter is to present a simple and transparent 
explanation for these questions.  It is crucial to examine an explicit 
form of the mode-coupling term on the semi-microscopic level beyond 
conventional group theoretical arguments which include ambiguities 
concerning combination of the wave vectors of order 
parameters~\cite{Kenzelmann,Aperis}.  The mode-coupling term relevant to 
the present situation, in the GL region, is given by the Feynman diagram 
shown in Fig.\ \ref{Fig:1}, and 
the analytic expression is given as follows: 
\begin{equation}
V_{\infty}=C\Delta^{(e)}_{\bf q}M_{{\bf Q}_0+{\bf q}}
\Delta^{(o)}_{-{\bf Q}_0},
\label{eq:1}
\end{equation}
where $M_{{\bf Q}_0+{\bf q}}$, $\Delta^{(e)}_{\bf q}$, and 
$\Delta^{(o)}_{-{\bf Q}_0}$ 
denote the IC-SDW magnetization in the $c$-direction, 
the $d$-wave superconducting (SC) gap in the 
FFLO state, and an additional odd-parity "equal-spin" SC gap with finite 
center-of-mass momentum $-{\bf Q}_{0}$, 
${\bf Q}_0\equiv(0.5\pi/a,0.5\pi/a,0.5\pi/c)$, (the so-called 
$\pi$-paring), respectively.  
The coefficient $C$ in eq.(\ref{eq:1}) is given in terms of the Green 
function $G_{\sigma}({\bf k},{\rm i}\epsilon_{n})$ of the quasiparticles 
in the normal state as follows: 
\begin{eqnarray}
& &C=g_{{\bf q}}g_{-Q_{0}+{\bf q}/2}T\sum_{n}\sum_{{\bf p}}\sum_{\sigma=\pm}
\phi_{\rm e}({\bf p})\chi^{(e)}_{\sigma{\bar \sigma}}
\phi_{\rm o}({\bf p}-{\bf Q}_{0}/2)\chi^{(o)}_{{\bar \sigma}{\bar \sigma}}
\times
\nonumber
\\
& &
\qquad
G_{\sigma}({\bf p}+{\bf q}/2,{\rm i}\epsilon_{n})
G_{{\bar \sigma}}({\bf p}-{\bf Q}_{0}-{\bf q}/2,{\rm i}\epsilon_{n})\times
\nonumber
\\
& &
\qquad
G_{{\bar \sigma}}(-{\bf p}+{\bf q}/2,-{\rm i}\epsilon_{n}),
\label{eq:2}
\end{eqnarray}
where $g$'s denote the pairing interactions for different center-of-mass 
momentum, and $\phi_{\rm e}({\bf p})$ is the wave function of $d$-wave 
pairing: e.g.,
$\phi_{\rm e}({\bf p})=(\cos\,p_{x}a-\cos\,p_{y}a)$, and 
$\phi_{\rm o}({\bf p})$ is the wave 
function of odd-parity state.  
$\chi^{(e)}_{\sigma{\bar \sigma}}$ and 
$\chi^{(o)}_{{\bar \sigma}{\bar \sigma}}$ 
denote the spin part of gap functions for even- 
and odd-party state, respectively.  Here, the quantization axis of 
"spin" (label of crystalline-electric-field ground doublet) is taken as 
parallel to the magnetic field, i.e., in the (1,1,0) direction, and the 
direction of IC-SDW magnetization is along the (0,0,1) direction.  
Since $\chi^{(e)}_{\sigma{\bar \sigma}}$  is antisymmetric with respect to 
interchange of $\sigma$ and ${\bar \sigma}$, i.e., 
$\chi^{(e)}_{{\bar \sigma}\sigma}=-\chi^{(e)}_{\sigma{\bar \sigma}}$, and 
$\chi^{(o)}_{{\bar \sigma}{\bar \sigma}}$ is not antisymmetric, 
the coefficient $C$ vanishes without the magnetic field: In other word, 
$C$ is proportional to the magnetic field $H$ or odd-function of $H$.  
As a result, the mode-coupling term eq.(\ref{eq:1}) 
becomes the time-reversal invariant quantity.

In order to explore the condition for the IC-SDW to be realized in practice, 
let us perform the analysis based on the GL free energy.  
For concise presentation, we denote $M_{{\bf Q}_0+{\bf q}}$ as ${\tilde M}$ 
and $\Delta^{(o)}_{-{\bf Q}_0}$ 
as ${\tilde \Psi}$.  Then, the GL free energy $F$ is expressed as 
\begin{eqnarray}
& &F=F_{0}+{a_{M}\over 2}{\tilde M}^{2}+{b_{M}\over 4}{\tilde M}^{4}+ 
{a_{\Psi}\over 2}{\tilde \Psi}^{2}+{b_{\Psi}\over 4}{\tilde \Psi}^{4}+
\nonumber
\\
& &\qquad\qquad
+C\Delta_{q}{\tilde M}{\tilde \Psi},
\label{eq:3}
\end{eqnarray}
where the coefficients $a_{M}$, $a_{\Psi}$, $b_{M}$, and $b_{\Psi}$ are 
assumed to be positive because both of ${\tilde M}$ and ${\tilde \Psi}$ 
are assumed to be vanishing without the FFLO gap $\Delta_{q}$.  The extremum 
conditions for ${\tilde M}$ and ${\tilde \Psi}$ are 
\begin{equation}
0={\partial F\over \partial {\tilde M}}=
a_{M}{\tilde M}+b_{M}{\tilde M}^{3}+C\Delta_{q}{\tilde \Psi},
\label{eq:4}
\end{equation}
and 
\begin{equation}
0={\partial F\over \partial {\tilde \Psi}}=
a_{\Psi}{\tilde \Psi}+b_{\Psi}{\tilde \Psi}^{3}+C\Delta_{q}{\tilde M}.
\label{eq:5}
\end{equation}
Considering that CeCoIn$_5$ is near the antiferromagnetic (AF) quantum 
critical point (QCP) with the ordering wave vector ${\bf Q}_{0}$ without 
the magnetic field so that the IC-SDW whose ordering vector 
${\bf Q}={\bf Q}_{0}+{\bf q}$ is also not far from the critical point, 
and anticipating the smallness of $\pi$-gap ${\tilde \Psi}$, 
we neglect the cubic term of ${\tilde \Psi}$ in eq.(\ref{eq:5}) to obtain 
\begin{equation}
{\tilde \Psi}=-{C\Delta_{q}\over a_{\Psi}}{\tilde M}.
\label{eq:6}
\end{equation}
Substituting this into eq.(\ref{eq:4}), we obtain the equation for 
${\tilde M}$: 
\begin{equation}
a_{M}{\tilde M}+b_{M}{\tilde M}^{3}-{(C\Delta_{q})^{2}\over a_{\Psi}}
{\tilde M}=0.
\label{eq:7}
\end{equation}
This has ${\tilde M}\not=0$ solution if the condition
\begin{equation}
(C\Delta_{q})^{2}>a_{\Psi}a_{M},
\label{eq:8}
\end{equation}
is satisfied, giving the IC-SDW magnetization as 
\begin{equation}
{\tilde M}^{2}={1\over b_{M}}\left[
{(C\Delta_{q})^{2}\over a_{\Psi}}-a_{M}\right]. 
\label{eq:9}
\end{equation}

The condition (\ref{eq:8}) will be satisfied if $a_M$ is small enough and 
$C\Delta_{q}$ is large enough.  The former condition may be guaranteed 
by some circumstantial evidence that CeCoIn$_5$ is located not far from 
the AF-QCP~\cite{Pham,Miyake}, and the latter condition is met in the 
FFLO state near the $H_{{\rm c}2}$ where the SC gap $\Delta_{q}$ remains 
to be large enough due to the first order transition at $H_{{\rm c}2}$.  
The rather large value of $H_{{\rm c}2}$ (comparable to the effective 
Fermi energy) supports the large value of the coefficient $C$.  

In the present model, the wave-vector of IC-SDW ${\bf Q}$ is not the same as 
that of FFLO ${\bf q}$, but is given by 
\begin{equation}
{\bf Q}={\bf Q}_{0}+{\bf q}. 
\label{eq:10}
\end{equation}
In the experiment of Ref.\ 1, the FFLO wave vector is considered to be 
parallel to $(1,1,0)$ so that the expected IC-SDW wave vector is 
along (0.5+q,0.5+q,0.5) direction in consistent with the experiment 
of ref.\ 1.  However, if the FFLO vector was parallel to $(1,0,0)$ 
or $(0,1,0)$, the expected IC-SDW wave vector would be in $(0.5+q,0.5,0.5)$ 
or $(0.5,0.5+q,0.5)$ direction, respectively.  This is a prediction of the 
present theory.  

If the magnitude of FFLO wave-vector $q$ is given by 
$q=2\mu_{\rm B}H/\hbar v_{\rm F}$, $Q$ has some dependence on the magnetic 
field $H$.  However, the relative effect is weaken by the presence of 
$Q_{0}$ in eq.(\ref{eq:10}) in contrast to the case where ${\bf Q}$ itself 
is identified with FFLO wave-vector as discussed in ref.\ 1.  Indeed, 
$q$ is estimated in the FFLO phase as follows: 
\begin{equation}
q={2\mu_{\rm B}H\over \hbar v_{\rm F}}\simeq 2.4\times10^{8}\,
({\rm m}^{-1}),
\label{eq:11}
\end{equation}
where we have used $H\simeq 10$ Tesla and 
$v_{\rm F}\simeq 7.5\times 10^{3}$ m~\cite{Miclea}.  Therefore, 
the ratio of $q$ and $Q_{0}$\ ($=a/\pi$) is 
\begin{equation}
{q\over Q_{0}}={4.6\times 10^{-10}\over \pi}\times 
2.4\times10^{8}\simeq 3.3\times 10^{-2}.
\label{eq:12}
\end{equation}
This value is about 3 times smaller than the observed value 
$(0.56-0.5)/0.5=0.12$.  However, the Fermi velocity $v_{\rm F}$ used 
for the estimation in eq.(\ref{eq:11}) is that near the SC transition 
temperature $T_{{\rm c}0}$ at $H=0$, and the mass enhancement at 
temperatures in FFLO phase is about 5-6 times larger than that 
at $T\sim T_{{\rm c}0}$ as estimated by the entropy balance argument.  
Therefore, it is reasonable that $q$, the deviation from the AF wave vector 
$Q_{0}$, is given by the wave-vector of FFLO state.  

The robustness of IC-SDW wave vector $Q$ against the magnetic field $H$ is 
understood as follows: The range of variation of $H$ is 
$10.6<H({\rm Tesla})<11.4$ 
so that the relative variation from the central value $H=11$ is about 
$3.6\times 10^{-2}$.  This means the absolute value of variation of 
$Q$ is $0.06\times 3.6\times 10^{-2}=2.1\times 10^{-3}$ which seems to 
be comparable to the error bar of IC-SDW wave vector although there 
is no error bar shown in the figure of Ref.\ \ref{Fig:1}.  Another 
factor, which reduces the effect of the magnetic field $H$, is the 
$H$-dependence of the Fermi velocity which is an increasing function 
of $H$ in general.  

The existence of the $\pi$-pairing of odd-parity can be detected, in 
principle, by the NQR measurement of the dynamical susceptibility along 
the magnetic field (in (1,1,0) direction), which can give rise to an 
anomalous relaxation of NQR spectrum as observed in 
Sr$_2$RuO$_4$~\cite{Mukuda} with the same tetragonal symmetry 
as in CeCoIn$_5$.  Indeed, since the wave function of 
$\phi_{\rm o}$ of the Cooper pair can be doubly degenerate in the 
odd-parity manifold, the spin-orbit coupling associated with relative 
motion of Cooper pairs can give rise to anomalous relaxation of 
NQR corresponding to the oscillations of magnetization along the 
(1,1,0) direction~\cite{KM}.  However, it may be left for future 
studies to analyze in more detail how to detect the $\pi$-pairing.  

In conclusion, the mechanism of IC-SDW order in the FFLO phase of CeCoIn$_5$ 
observed by neutron scattering has been clarified on the basis of a new 
mode-coupling scheme among IC-SDW order, 
two SC orders of FFLO with B$_{1{\rm g}}$ 
($d_{x^{2}-y^{2}}$) symmetry and $\pi$-pairing with odd-parity.  
The mode-coupling term proposed in the present Letter gives a simple 
explanation for why the IC-SDW order is observed only in FFLO phase 
and the IC wave vector is rather robust against the magnetic field.  
The difference between the case of CeRhIn$_5$~\cite{Knebel}, 
which is a sister compound of CeCoIn$_5$ and IC-SDW state at $H=0$, 
should be clarified by further investigations.  

\section*{Acknowledgements}
The author acknowledges the hospitality of SPSMS, CEA/Grenoble 
where this work was performed.  The author benefited from conversations 
and clarifying discussions with D. Aoki, J. Flouquet, and G. Knebel 
who took my attention to this 
problem.  
This work is supported by a Grant-in-Aid for Specially Promoted Research 
(20001004) from the Ministry of Education, Culture, Sports, Science 
and Technology.

\end{document}